\documentclass[12pt,a4paper,final]{iopart}
\usepackage{cite}
\usepackage{bm}
\usepackage{float}
\usepackage{stmaryrd,scalerel} 
\usepackage{iopams}
\usepackage{color}
\usepackage[breaklinks=true,colorlinks=true,linkcolor=blue,urlcolor=blue,citecolor=blue]{hyperref}
\expandafter\let\csname equation*\endcsname=\relax
\expandafter\let\csname endequation*\endcsname=\relax
\usepackage{amsmath}

\usepackage{url}

\expandafter\let\csname equation*\endcsname\relax

\expandafter\let\csname endequation*\endcsname\relax

\usepackage{amsmath}

\begin{document}

\title{Diffusing diffusivity model with dichotomous noise}
\author{Dongho Lee$^1$, Jae-Hyung Jeon$^{1,2}$, Pascal Viot$^3$ \& Gleb Oshanin$^{2,3}$}
\address{$^1$Department of Physics, Pohang University of Science and Technology (POSTECH), Pohang 37673, Republic of Korea\\
$^2$Asia Pacific Center for Theoretical Physics (APCTP),
Pohang 37673, Republic of Korea\\
$^3$Sorbonne Universit\'e, CNRS, Laboratoire de Physique Th\'eorique de la Mati\`ere Condens\'ee (UMR CNRS 7600), 4 Place Jussieu, 75252 Paris Cedex 05, France }

\date{today}

\begin{abstract}
	We study Langevin dynamics with stochastic diffusivity arising from fluctuations of the surrounding medium. The diffusivity is modeled as Ornstein–Uhlenbeck process driven by symmetric dichotomous noise, which confines it to a finite interval. We derive analytical expressions for the short-time probability density function (PDF) of the particle displacement and analyse its asymptotic behaviour. While the PDF retains the characteristic logarithmic divergence at the origin, its tails differ from the Gaussian white-noise case: exponential tails are replaced by Gaussian ones modulated by a power-law with a switching-rate-dependent exponent. At long times, the dynamics converges to ordinary Gaussian diffusion. We determine the variance and covariance of the time-averaged stochastic diffusivity and show that it is self-averaging. The model provides a minimal analytically tractable framework for stochastic transport in environments with bounded or switching fluctuations.
		\end{abstract}
		
	
	\section{Introduction}
	
Understanding stochastic transport in heterogeneous media remains a central problem across many fields. In the classical description of Brownian motion, the mean-square displacement (MSD) of a tracer particle grows linearly in time, and the displacement probability density function (PDF) is Gaussian. However, numerous experiments in complex systems have demonstrated that, while the MSD retains this linear scaling, the displacement PDF may exhibit pronounced non-Gaussian features~\cite{wangAnomalousBrownian2009a,jeonProteinCrowdingLipid2016a}. This phenomenon, commonly referred to as Brownian yet non-Gaussian diffusion, has been observed in a wide range of systems, including crowded colloidal suspensions~\cite{guanEvenHardSphereColloidal2014,matseTestDiffusingdiffusivityMechanism2017a,ciarloFickianNonGaussianDiffusion2023}, biological environments such as actin networks and cellular membranes~\cite{wangAnomalousBrownian2009a,slezakDiffusionCompartmentalizedMedia2021a}, and glassy or dynamically heterogeneous soft-matter systems~\cite{brizioliReciprocalSpaceStudy2022a,kwonDynamicsCrowdedEnvironments2014}. A recent review~\cite{akimotoAnomalousStatisticsLangevin2026} provides a comprehensive overview of experimental evidence for deviations from Gaussian behavior, together with a systematic presentation of theoretical frameworks, including diffusing diffusivity models.
Taken together, these observations indicate that even when the MSD exhibits normal diffusive scaling, the underlying dynamics may be governed by temporal or spatial fluctuations of the local environment, leading to significant deviations from Gaussian displacement statistics.

A widely used theoretical framework for describing such behaviour is the diffusing diffusivity (DD) model introduced by Chubynsky and Slater~\cite{chubynskyDiffusingDiffusivityModel2014a}. In this approach, the diffusivity itself is treated as a stochastic process representing environmental fluctuations experienced by the tracer particle. In the minimal model introduced by Chechkin \textit{et al.}~\cite{chechkinBrownianNonGaussianDiffusion2017} and independently by Lanoisel\'ee and Grebenkov \cite{lanoiseleeModelNonGaussianDiffusion2018a} (see also~\cite{sposiniRandomDiffusivityStochastic2018a,lanoiseleeDiffusionlimitedReactionsDynamic2018a,sposiniUniversalSpectralFeatures2020a}), the diffusivity is defined as the square of Ornstein–Uhlenbeck (OU) process driven by Gaussian white noise; we refer to this formulation as the Gaussian OU case. This model reproduces the coexistence of a linear MSD and non-Gaussian displacement distributions at intermediate times, while Gaussian statistics are recovered asymptotically. Extensions of this framework, including studies of extremal statistics and first-passage properties, have been reported in~\cite{jainDiffusingDiffusivityFractional2018a,grebenkovExactFirstpassageTime2021,grebenkovExactDistributionsMaximum2021a,hidalgo-soriaCuspNonGaussianDensity2021a,deanExactHeightDistribution2025a}.

The formulations proposed in \cite{chechkinBrownianNonGaussianDiffusion2017,lanoiseleeModelNonGaussianDiffusion2018a} implicitly assume that fluctuations of the local environment are continuous and unbounded. In many physical systems, however, environmental variability is better described by transitions between a finite set of dynamical states occurring at random times. Representative examples include transport in compartmentalized media \cite{weronErgodicityBreakingNeuronal2017a}, systems with switching activity \cite{bressloffStochasticSwitchingBiology2017a,songMachineLearningApproach2023a,parkSingleTrajectoryBayesianModeling2025a}, and heterogeneous materials in which local mobility alternates between metastable configurations \cite{hachiyaUnveilingDiffusiveStates2019a,pastoreDynamicPhaseCoexistence2015a}. In such settings, the effective diffusivity evolves via intermittent switching between discrete dynamical regimes rather than through continuous fluctuations. These processes have been extensively studied within the framework of switching diffusion models, where the diffusivity randomly alternates among a finite set of states~\cite{miyaguchiLangevinEquationFluctuating2016,grebenkovUnifyingApproachFirstpassage2019}. Closely related descriptions arise in stochastic dynamics driven by dichotomous (telegraph) noise, which provides a minimal model for fluctuations between two metastable configurations (see, e.g.,~\cite{horsthemkeNoiseInducedTransitionsTheory2006,haunggiColoredNoiseDynamical1994,benaDichotomousMarkovNoise2006a,gardinerStochasticMethodsHandbook2009a,umLangevinDynamicsDriven2019a}).
This motivates generalized DD models in which the stochastic driving remains bounded and undergoes random switching between a finite set of discrete states.

In this paper, we consider a modification of the minimal DD model in which the OU process governing the stochastic diffusivity is driven by symmetric dichotomous (telegraph) noise instead of Gaussian white noise. The mean-reverting structure of the OU dynamics is preserved, while the stochastic forcing switches randomly between two discrete values only. As a result, the stationary diffusivity is confined to a finite interval. This formulation, hereafter referred to as the dichotomous OU case, provides a minimal and analytically tractable framework for stochastic transport under bounded environmental fluctuations. We show that this modification leads to quantitative differences in the displacement statistics : At short times, the position PDF can be expressed in terms of the confluent hypergeometric (Tricomi) function. The behaviour near the origin remains logarithmically divergent, reflecting the same small-diffusivity structure as in the Gaussian OU model. In contrast, the distribution tails differ qualitatively: in the Gaussian OU case they decay exponentially, whereas in the dichotomous OU case they are Gaussian, divided by a power-law dependence on the position with a non-universal exponent. This difference originates from the bounded support of the diffusivity and from the switching-rate-dependent form of its stationary distribution. At long times, both models converge to Gaussian diffusion.

The remainder of the paper is organized as follows. In Sec.~\ref{model} we briefly recall the DD framework introduced in~\cite{chechkinBrownianNonGaussianDiffusion2017,lanoiseleeModelNonGaussianDiffusion2018a} and define the dichotomous OU case to be considered here. In Sec.~\ref{PDF} we analyse the time evolution of the displacement statistics in the short- and long-time regimes, derive analytical expressions for the probability density function, and compare them with the predictions of the Gaussian OU model. Finally, Sec.~\ref{concs} summarizes the results and outlines some perspectives for further research. Details of the numerical simulations are provided in ~\ref{A}. The derivation of the variance and covariance of the time-averaged diffusivity driven by dichotomous noise is presented in ~\ref{D}.

\section{Model}
\label{model}

We begin by briefly recalling the framework of \cite{chechkinBrownianNonGaussianDiffusion2017,lanoiseleeModelNonGaussianDiffusion2018a}. Consider a particle performing random motion on an infinite one-dimensional line and assume that its instantaneous position 
$X(t)$ evolves according to the Langevin equation
	\begin{equation}
		\label{LE}
		\frac{\mathrm{d} X(t)}{\mathrm{d} t} = \sqrt{2 D_t} \, \zeta_t , \quad X(0) = 0 \,, \quad 0 \leq t < \infty \,, 
	\end{equation}
	where $\zeta_t$ is a Gaussian white noise process with zero mean and the covariance function
	\begin{equation}
		\overline{\zeta_t \, \zeta_{t'}} =   \delta(t-t') \,.
		\end{equation}
		In the above equation and henceforth the overbar denotes averaging over all possible realizations of the thermal noise and $\delta(t)$ is the Dirac delta-function.
	
In Eq.~\eqref{LE}, 
$D_t$
represents a stochastic diffusivity. The key assumption of the DD model introduced in \cite{chechkinBrownianNonGaussianDiffusion2017,lanoiseleeModelNonGaussianDiffusion2018a} is that fluctuations of the transport coefficient originate from an underlying dynamical environment and can therefore be modeled by an auxiliary stochastic process. Specifically, in \cite{chechkinBrownianNonGaussianDiffusion2017,lanoiseleeModelNonGaussianDiffusion2018a} the diffusivity is expressed as 
\begin{align}
	\label{b}
D_t=Y_t^2 \,, \quad D(t) \geq 0 \,,
\end{align}
where the auxiliary real-valued function 
$Y_t$ (physical dimension is length/time$^{1/2}$)
is the OU process; that being, $Y_t$ obeys 
\begin{align}
	\label{c}
		\frac{\mathrm{d}Y_t}{\mathrm{d}t} = -\frac{1}{\tau} Y_t  + \sigma \, \eta_t, \quad -\infty < t < \infty \,,
	\end{align} 
with $\tau$ being the relaxation time, $\sigma$ - the noise amplitude (dimension length/time), and $\eta_t$ - a  Gaussian white  noise process (dimension 1/time$^{1/2}$) with zero mean and the covariance function
\begin{align}
	\label{cov}
	\langle \eta_t \, \eta_{t'} \rangle_{\eta} =   \delta(t-t') \,.
	\end{align}
	The angle brackets with the subscript $\eta$ here and henceforth denote averaging over realizations of noise $\eta_t$. 
	Before proceeding, we note that 
	$D(t)$ in such setting
	corresponds, in fact, to a Cox–Ingersoll–Ross (CIR) process\cite{coxTheoryTermStructure1985}. Specifically, by applying It\^{o}’s formula to Eq.~\eqref{c}, one finds that the squared OU variable 
	$D_t=Y_t^2$
	naturally satisfies a CIR-type stochastic differential equation \cite{coxTheoryTermStructure1985} (see also \cite{chechkinBrownianNonGaussianDiffusion2017,lanoiseleeModelNonGaussianDiffusion2018a}),
\begin{equation}
	\frac{\mathrm{d}D_t}{\mathrm{d}t} = \frac{2}{\tau} \left(\frac{\sigma^2 \tau}{2} - D_t \right) + 2\sigma \sqrt{D_t}\, \eta_t \,,
\end{equation}	
	 in which the diffusivity remains positive and exhibits mean-reverting behavior. Note that this choice of stochastic dynamics has previously been used in the mathematical finance literature by Stein and Stein \cite{steinStockPriceDistributions1991} and by Heston \cite{hestonClosedFormSolutionOptions1993} to introduce a stochastic evolution of volatility in generalizations of the Black–Scholes model\cite{blackValuationOptionContracts1972, blackPricingOptionsCorporate1973} (see also~\cite{bouchaudTheoryFinancialRisk2003} for a broader perspective and critical assessment). In the present context, it provides a natural mechanism for time-dependent particle diffusivity, ensuring physical consistency and generating the non-Gaussian displacement distributions observed at intermediate times in the DD framework in \cite{chechkinBrownianNonGaussianDiffusion2017,lanoiseleeModelNonGaussianDiffusion2018a}.

In this paper, as an alternative to the standard Gaussian OU model, we consider a discrete-valued symmetric dichotomous noise to drive 
$Y_t$. Specifically, we define 
$Y_t$
through the stochastic differential equation
	\begin{equation}
		\label{dicho}
	\dot Y_t = - \frac{1}{\tau} Y_t + A \, \xi_t \,, \quad -\infty < t < \infty \,, 
\end{equation}
where $\xi_t$ is a symmetric dichotomous noise  that switches randomly between two values
\begin{align}
	\xi_t = \pm 1 \,.
	\end{align}
The transitions between the two states are memoryless, and the time intervals between successive switching events are independent, exponentially distributed random variables. The respective switching rate is denoted by $\lambda$. Consequently, the mean and the autocorrelation of the noises obey \cite{horsthemkeNoiseInducedTransitionsTheory2006,haunggiColoredNoiseDynamical1994,benaDichotomousMarkovNoise2006a,gardinerStochasticMethodsHandbook2009a}
\begin{align}
	\langle	\xi_t  \rangle_{\xi} = 0 \,, \quad \langle \xi_t \xi_{t'} \rangle_{\xi} = \exp\left(-2 \lambda |t - t'|\right) \,,
	\end{align}
where the angle brackets with the subscript $\xi$ denote here and henceforth averaging with respect to the dichotomous noise. 
In turn, the parameter $A$ in Eq. \eqref{dicho}
sets the amplitude of the fluctuations, while 
$\tau$ (as above) is the relaxation time. The amplitude 
$A$ has physical dimension 
length/time$^{3/2}$, ensuring that the derived process 
$D_t$
retains the correct units of diffusivity. 

As  remarked above, the above dichotomous OU model is motivated by systems in which fluctuations are naturally finite and discrete, such as molecular switches, active forces in complex media, or telegraphic-like environmental perturbations \cite{horsthemkeNoiseInducedTransitionsTheory2006,haunggiColoredNoiseDynamical1994,benaDichotomousMarkovNoise2006a,gardinerStochasticMethodsHandbook2009a,umLangevinDynamicsDriven2019a}. Alternatively, 
$Y_t$
admits a physical interpretation as the instantaneous position of a particle evolving under run-and-tumble dynamics in a quadratic confining potential \cite{tailleurStatisticalMechanicsInteracting2008a,gueneauActiveParticleHarmonic2023}.
In the stationary regime, the process 
$Y_t$ is confined to the finite interval 
$[- A \tau, A \tau]$. Its stationary distribution continuously interpolates between a 
$U$-shaped form in the rare-switching limit and a quasi-Gaussian shape in the fast-switching limit \cite{sanchoStochasticProcessesDriven1984} (see also the recent discussion in \cite{herbeauStochasticGyrationDriven2026}). This behavior is in sharp contrast with the Gaussian OU process in Eq.~\eqref{c}, whose stationary distribution is Gaussian with unbounded support. As a direct consequence of the boundedness of 
$Y_t$, the diffusivity 
$D_t=Y_t^2$
in the dichotomous OU case is supported on the finite interval 
$[0,(A \tau)^2]$ and follows a beta distribution (see below) in the stationary state.

We finally note that the dichotomous driving term 
$A \, \xi_t$ in Eq. \eqref{dicho}
 reduces to Gaussian white noise in the right-hand-side in Eq. \eqref{c}  in a suitable scaling limit. Specifically, consider the double limit
\begin{align}
	\label{limit}
	A \to \infty \,, \quad \lambda \to \infty \,,
	\end{align}
while keeping the ratio 
$A^2/\lambda=\sigma^2$
fixed. In this regime, the correlation time 
$1/(2 \lambda)$
vanishes, whereas the noise intensity remains finite. Under these conditions, the process 
$A \, \xi_t$
converges, in the sense of weak convergence, to Gaussian white noise of the form 
$\sigma \, \eta_t$, where 
$\eta_t$ is a normalized white noise with covariance in Eq. \eqref{cov} \cite{horsthemkeNoiseInducedTransitionsTheory2006,haunggiColoredNoiseDynamical1994,benaDichotomousMarkovNoise2006a,gardinerStochasticMethodsHandbook2009a}. This limiting procedure will serve for us as a consistency check: in the limit of fast switching of the dichotomous noise, the dichotomous OU dynamics should reduce to the effective Gaussian OU process that defines the DD model introduced in \cite{chechkinBrownianNonGaussianDiffusion2017,lanoiseleeModelNonGaussianDiffusion2018a}.

\section{Temporal evolution of the position probability density function}
\label{PDF}

We now turn to the temporal evolution of the position probability density function 
$P(X,t)$
 for the two models. As a first step, it is instructive to compare the behavior of the second moment 
$\overline{\langle X_t^2\rangle}$, which provides a direct measure of the effective spreading rate. 
In both cases -- the Gaussian OU and the dichotomous  OU -- the mean value of $X_t$ vanishes by symmetry, while the second moment obeys
\begin{align}
	\label{z}
	\frac{\mathrm{d}}{\mathrm{d} t} \overline{\langle X_t^2\rangle} = 2 \langle D_t\rangle \,,
	\end{align}
	being controlled entirely by the mean of the diffusivity process. 
	In both models $Y_t$ is 
	mean-reverting and reaches  a stationary state with finite variance,  implying that $\langle D_t\rangle$ converges to a constant. Consequently, in both models the MSD exhibits linear growth, $\overline{\langle X_t^2\rangle} = 2 D_{\rm eff} t$,  where $D_{\rm eff} = \sigma^2 \tau/2$ for the Gaussian OU case \cite{chechkinBrownianNonGaussianDiffusion2017,lanoiseleeModelNonGaussianDiffusion2018a}, while in our case of the dichotomous OU process we find
	 \begin{align}
	 	\label{z2}
	 	D_{\rm eff} = \frac{A^2 \tau^2}{1 + 2 \lambda \tau} \,,
	 	\end{align}
	 	showing explicitly how the switching rate 
	 	$\lambda$ modulates the effective transport. We note that in the diffusive limit (see Eq. \eqref{limit}) the result in Eq. \eqref{z2} coincides with the above expression for the diffusion coefficient in the Gaussian OU case, as it should.

Thus, while both models produce normal diffusion at long times, they differ quantitatively in their effective diffusion coefficients and, as we shall see below, more substantially in the full time-dependent shape of the probability density 
$P(X,t)$, particularly at short times where non-Gaussian features are most pronounced.	We present such an analysis below.

	\subsection{Short-time evolution of $P(X,t)$}

We assumed that both stochastic diffusivity processes (Gaussian OU and dichotomous OU) are started at 
$t = - \infty$, so that 
$D_t=Y_t^2$
is already in its stationary state at time 
$t=0$. This is crucial: it eliminates transient effects in 
$D_t$
and isolates the intrinsic effect of diffusivity fluctuations on the particle motion.

For times shorter than the correlation time $\tau_c$ of 
$D_t$, 
where 
$\tau_c=\tau$ for the Gaussian OU and 
$\tau_c= \min(\tau,1/\lambda)$
 for the dichotomous OU process,
the diffusivity does not significantly change during a trajectory segment. It is effectively quenched.
In virtue of Eq. \eqref{LE}, the position PDF $P(x,t|D)$ conditioned on a fixed value of 
$D$ is Gaussian:

\begin{align}
P(X,t|D) = \frac{1}{\sqrt{4\pi D t}} \,\exp\left(-\frac{X^2}{4 D t}\right).
\end{align}

Since 
$D$ is itself random (drawn from its stationary distribution 
$P_{\rm st}(D)$)
the unconditional PDF is a superposition of Gaussians with fluctuating variances:
\begin{equation}
	\label{m}
	P(X,t)  = \int^{\infty}_0  \mathrm{d}D \, P(X,t|D)  \, P_{\rm st}(D) \,.
\end{equation}
Note that this regime is the hallmark of superstatistics \cite{beckSuperstatistics2003a}, in which a fast process (here, Brownian motion) is driven by a slowly fluctuating parameter $D$. The total statistics is obtained by averaging over the stationary distribution of that parameter.

For the Gaussian OU model the stationary distribution $P_{\rm st}(D)$ is the gamma-distribution with the shape $1/2$ \cite{chechkinBrownianNonGaussianDiffusion2017,lanoiseleeModelNonGaussianDiffusion2018a}:
\begin{align}
	P_{\rm st}(D) = \frac{1}{\sqrt{\pi \sigma^2 \tau}} \, D^{-1/2} \,\exp\left(- \frac{D}{\sigma^2 \tau}\right),
\end{align}	
with $D$ being defined on the entire half-line, $D \in [0,\infty)$. Respectively, the unconditional position PDF is
\begin{align}
	\label{K0}
	P(X,t) =   \frac{1}{\pi \sqrt{\sigma^2 \tau t }} \, K_0\left(\frac{|X|}{\sqrt{\sigma^2 \tau t}}  \right) \,, 
	\end{align}
where $K_0(z)$ is the modified Bessel function of the second kind \cite{batemanHigherTranscendentalFunctions1955}. The appearance of this function encodes two striking non-Gaussian features: First, for small values of the argument, one has $K_0(z) \sim -\ln z$, so that the PDF exhibits a logarithmic divergence at the origin, reflecting the strong contribution of trajectories with very small instantaneous diffusivity. Second, for large values of the argument, $K_0(z)\sim \mathrm{e}^{-z}$, yielding exponential spatial tails rather than Gaussian ones. These non-Gaussian features coexist with a strictly linear MSD, illustrating how fluctuating diffusivity can generate deviations from Gaussian statistics while preserving normal diffusion at the level of the MSD.

We turn next to the model under study in which $Y_t$ is driven by the dichotomous OU process. The probability density function of $Y_t$ has been studied in \cite{sanchoStochasticProcessesDriven1984} (see also recent \cite{herbeauStochasticGyrationDriven2026})
and its stationary form has also been determined. Changing the variable, $D = Y^2$, we read of from the result of \cite{sanchoStochasticProcessesDriven1984} the following exact expression for the stationary distribution function of the DD:

\begin{align}\label{eq:sanchobis}
	P_{\rm st}(D) = C \, D^{-1/2}
	 (A^2 \tau^2 - D)^{\lambda \tau - 1}, \qquad
	C = \frac{\Gamma(\lambda \tau+1/2) }{\sqrt{\pi} \Gamma(\lambda \tau) (A \tau)^{2 \lambda \tau-1}}, 
	\end{align}
where $D$ is now defined on a finite support, $D \in [0, (A \tau)^2]$, in contrast to the above considered Gaussian OU case. One observes that the behaviour of the stationary distribution near the right edge of its support depends sensitively on the value of the dimensionless parameter $\lambda \tau$, which for fixed $\tau$ is controlled by the switching rate $\lambda$ of the dichotomous noise. Specifically, three qualitatively distinct regimes emerge: When $\lambda \tau > 1$, the stationary probability density function $P_{\rm st}(D)$ smoothly approaches zero as $D$ tends to the maximal value $(A \tau)^2$, so that the right edge of the support is regular. In contrast, when $\lambda \tau < 1$, the distribution develops a power-law divergence as $D \to (A \tau)^2$. In this regime the stationary density acquires a pronounced $U$-shaped form, characterised by an integrable singularity at the upper boundary of the support. In the marginal case $\lambda \tau = 1$, the behavior becomes intermediate: the stationary probability density approaches a finite value as $D \to (A \tau)^2$. Independently of the value of $\lambda \tau$, the distribution exhibits a universal (integrable) singular behaviour at the opposite edge of the support. As $D \to 0$, the stationary density diverges as $P_{\rm st}(D) \propto D^{-1/2}$. This small-$D$ behavior coincides with that found in the DD model driven by Gaussian white noise \cite{chechkinBrownianNonGaussianDiffusion2017,lanoiseleeModelNonGaussianDiffusion2018a}, reflecting the generic accumulation of probability weight near vanishing instantaneous diffusivities.

\begin{figure}[t]
\includegraphics[width=0.31\textwidth]{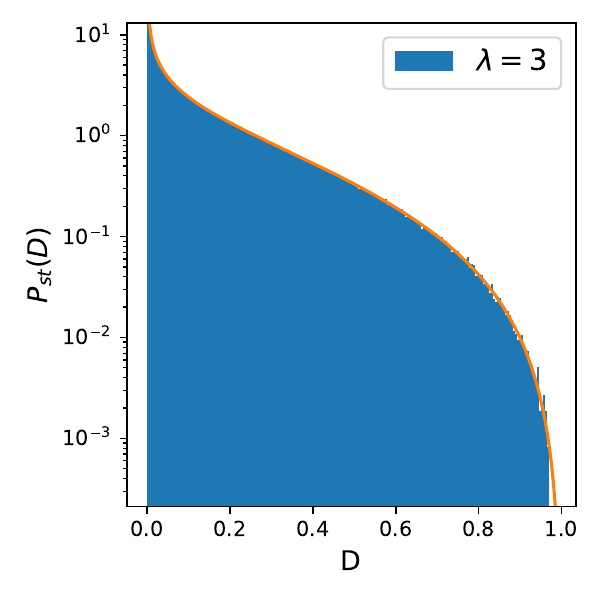}
\includegraphics[width=0.31\textwidth]{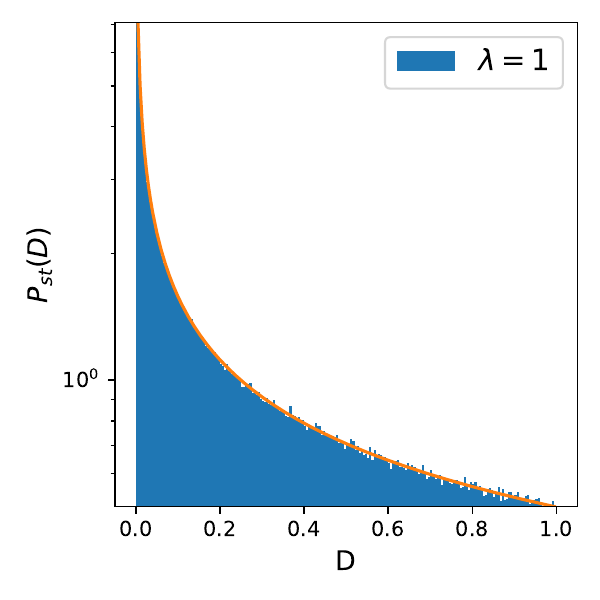}
\includegraphics[width=0.31\textwidth]{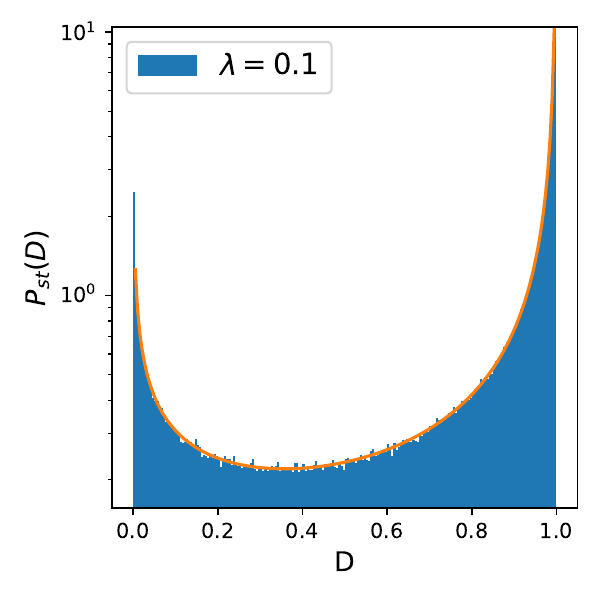}		\caption{Stationary probability density function 
		$P_{\rm st}(D)$
		 for 
		$\tau =1$, $A =  1$
		 and three values of the switching rate $\lambda$: 
		$\lambda =3$ (left), 
			$\lambda =1$ (center), and 
		$\lambda =0.1$ (right). The thin magenta lines represent the analytical predictions from Eq.~\eqref{eq:sanchobis}, while the blue shaded regions show histograms obtained from numerical simulations  (see ~\ref{A} for more details).}\label{fig:1}
\end{figure}

The theoretically predicted stationary distribution $P_{\rm st}(D)$ in the different regimes described above is shown in Fig.~\ref{fig:1}, where it is compared with the results obtained from numerical simulations. This Figure highlights the three distinct stationary distributions arising for different switching regimes and demonstrates excellent agreement between theory and simulations across the full range of $D$.

The expression \eqref{eq:sanchobis} entails the following general results for the moments of $D$ of arbitrary, not necessarily integer order $k$ :
\begin{align}
	\label{mom}
	\langle D^k\rangle_{\xi} = \frac{\Gamma(\lambda \tau + 1/2) \Gamma(k+1/2)}{\Gamma(k+\lambda \tau +1/2) \sqrt{\pi}} \left(A \tau\right)^{2 k} \,,
\end{align}
and for the averaged logarithm of $D$ :
\begin{align}
		\langle \ln D\rangle_{\xi} = 2 \ln\left(\frac{A \tau}{2}\right) - \gamma - \psi(1/2 + \lambda \tau) \,,
	\end{align}
	where $\gamma \approx 0.577$ is the Euler-Mascheroni constant, while $\psi$ is the digamma function \cite{batemanHigherTranscendentalFunctions1955}. The latter expression gives the so-called "typical" value of $D$, which should be observed for most of realizations of the dichotomous process, and  is formally 
	defined as
	\begin{align}
		D_{\rm typ} = \exp\left(\langle \ln D\rangle_{\xi}\right) \,.
		\end{align}
We find then that 
\begin{align}
	\frac{D_{\rm typ}}{\langle D\rangle_{\xi}} = \frac{(1 + 2 \lambda \tau)}{4} \mathrm{e}^{-\gamma - \psi(1/2 + \lambda \tau)} \,,
	\end{align}
	where $\langle D\rangle_{\xi}$ is the first moment of $D$ (see Eqs. \eqref{mom} and 
	\eqref{z2}). The function appearing in the right-hand side of the latter expression depends solely on the dimensionless parameter 
	$\lambda \tau$. It is a monotonically decreasing function of this parameter, approaching unity in the limit 
	$\lambda \tau \to 0$ and tending to $\exp(-\gamma)/2$
	 as $\lambda \tau \to \infty$. Importantly, this function remains strictly smaller than unity for any finite value of 
	$\lambda \tau$. This observation implies that the typical values of the diffusivity 
	$D$ are systematically smaller than its ensemble average. In other words, the mean value 
	$\langle D\rangle_{\xi}$, as well as higher moments of the diffusivity distribution, are dominated by relatively rare realizations of the dichotomous noise that produce unusually large values of 
	$D$.

We focus next on the short-time evolution of the position probability density function $P(X,t)$.
Inserting the expression \eqref{eq:sanchobis} into Eq. \eqref{m} and integrating, we find
\begin{align}
	\label{mm}
	\begin{split}
	P(X,t)  &= \frac{C}{\sqrt{4 \pi t}} \int^{A^2 \tau^2}_0  \frac{\mathrm{d}D}{D} \, \left(A^2 \tau^2 - D\right)^{\lambda \tau - 1}  \, \exp\left(- \frac{X^2}{4 D t}\right) \\
	&= \frac{\Gamma(\lambda \tau+1/2)}{2 \pi A \tau \sqrt{t}} \exp\left(- \frac{X^2}{4 A^2 \tau^2 t}\right) U\left(\lambda \tau, 1, \frac{X^2}{4 A^2 \tau^2 t}\right) \,,
	\end{split}
\end{align}
where $U$ is the confluent hypergeometric (Tricomi) function \cite{batemanHigherTranscendentalFunctions1955}. 

We now examine the expression in Eq. \eqref{mm}. 
For small values of its argument, the Tricomi confluent hypergeometric function admits an asymptotic expansion \cite{batemanHigherTranscendentalFunctions1955}

\begin{align}
	U\left(\lambda \tau, 1, \frac{X^2}{4 A^2 \tau^2 t}\right)  = - \frac{1}{\Gamma(\lambda \tau)}
 \left(
 \ln\left(
 \frac{X^2}{4 A^2 \tau^2 t}\right)   + 2 \gamma + \psi\left(\lambda \tau\right)
 \right) +O\left(\frac{X^2}{4 A^2 \tau^2 t}\right) \,.
 \end{align}
The above expansion 
   reveals a logarithmic divergence of the probability density function at the origin,
\begin{align}
	P(X,t) \simeq - \frac{\Gamma(\lambda \tau+1/2)}{2 \pi \Gamma(\lambda \tau) A \tau \sqrt{t}} \left( \ln\left(
	\frac{X^2}{4 A^2 \tau^2 t}\right) + 2 \gamma + \psi(\lambda \tau) \right) \,.
	\end{align}
This behavior is fully analogous to that found in the Gaussian OU case.
Note, that the same conclusion can, in fact, be reached directly from the integral representation of 
$P(X,t)$
given in the first line of Eq. \eqref{mm}. Setting 
$X=0$, one observes that the integral diverges logarithmically at the lower limit of integration. The logarithmic singularity at the origin therefore appears as a robust feature of both the Gaussian OU and the dichotomous OU dynamics and can be traced back to essentially identical statistics of the zero crossings of the underlying process 
$Y_t$ in the two cases.

Consider next the large-$X$ behavior of the PDF $P(X,t)$ in Eq. \eqref{mm}. Taking into account that $U(a,1,z) \simeq 1/z^{a}$ when $z \to \infty$ \cite{batemanHigherTranscendentalFunctions1955}, we get
\begin{align}
	P(X,t) \simeq \frac{\Gamma(\lambda \tau+1/2)}{2 \pi A \tau \sqrt{t}} \left(4 A^2 \tau^2 t\right)^{\lambda \tau} \exp\left(- \frac{X^2}{4 A^2 \tau^2 t}\right)/X^{2 \lambda \tau} \,.
	\end{align}
Therefore, unlike the Gaussian OU model, where the displacement PDF has exponential tails, the present model produces Gaussian tails, and moreover, with an additional algebraic suppression governed by a non-universal exponent set by the product of the switching rate $\lambda$ and the relaxation time $\tau$. This implies that for finite values of the noise amplitude, the dichotomous driving results in a position PDF that is more narrow and more strongly concentrated around the origin, as intuitively expected for fluctuations of bounded magnitude.

Lastly, to verify the internal consistency of the model, we consider the diffusive limit defined in Eq. \eqref{limit}. Employing the appropriate asymptotic relations for the special functions involved, 
\begin{align}
	\lim_{a \to \infty} \frac{\Gamma(a+1/2)}{a^{1/2} \Gamma(a)} = 1 \,, \quad \lim_{a \to \infty} \Gamma(a) \, U\left(a,1,\frac{z^2}{a}\right) = 2 K_0(2 |z|) \,,
	\end{align}
we find that the expression in Eq. \eqref{mm} reduces exactly to the form given in Eq. \eqref{K0}. This recovery of the known result confirms that, in the diffusive limit, the dichotomous OU model converges to the standard Gaussian OU one, as expected.

\begin{figure}[t]
	\includegraphics[width=0.31\textwidth]{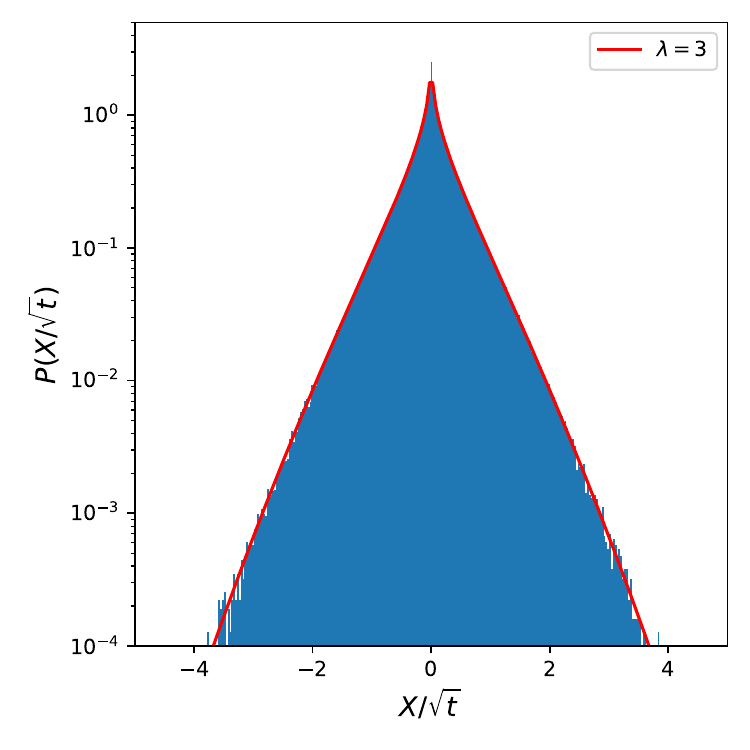}
\includegraphics[width=0.31\textwidth]{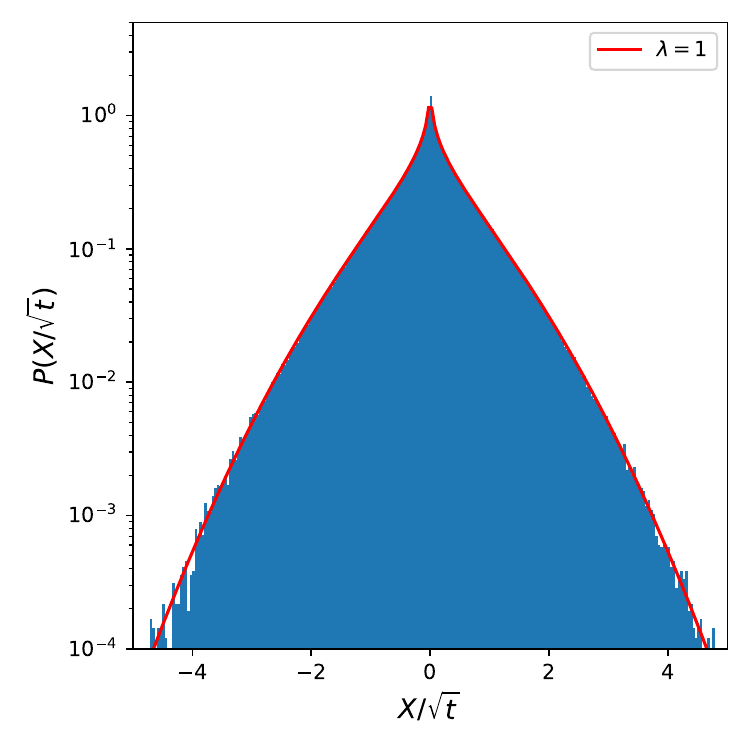}
\includegraphics[width=0.31\textwidth]{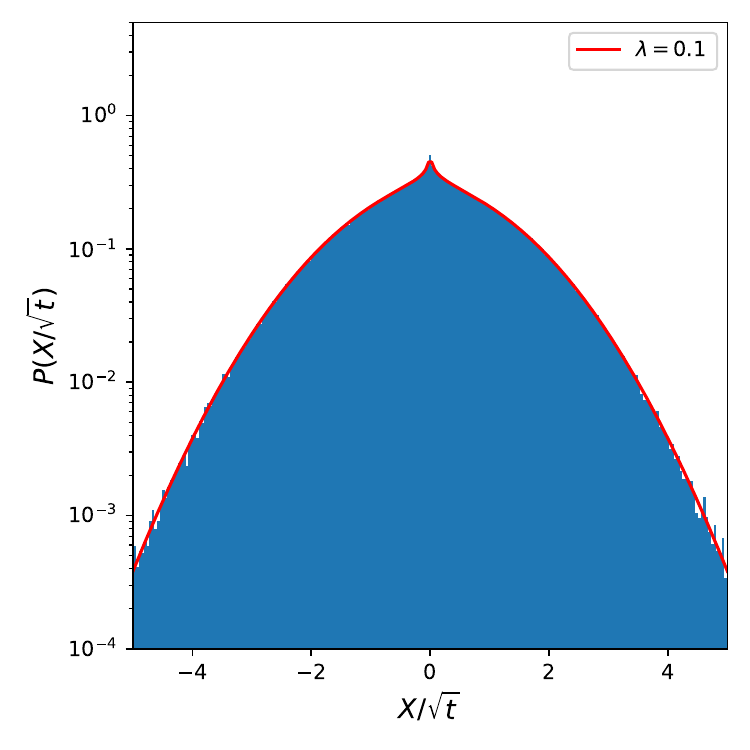}
		\caption{Short-time evolution of $P(X,t)$ in the DD model driven by dichotomous noise for $\tau = 1$, $A  = 1$ and three values of the switching rate  $\lambda$:
		$\lambda=3$ (left), $\lambda = 1$ (center) and $\lambda = 0.1$ (right). Thin red curves correspond to our theoretical prediction in Eq. \eqref{mm}. The blue histograms depict our numerical simulations results.}\label{fig:DPdeX}
\end{figure}

Figure~\ref{fig:DPdeX} illustrates the short-time behaviour of the position PDF, plotted as a function of the rescaled variable 
$X/\sqrt{t}$, for several values of the switching rate 
$\lambda$ (with the relaxation time fixed to 
$\tau = 1$  and the amplitude $A = 1$.). The numerical results (blue histograms) were obtained by sampling Eq.~\eqref{eq:sanchobis} using a rejection-free simulation algorithm (see \ref{A} for details). The thin red curves represent the analytical prediction given in Eq.~\eqref{mm}.
The simulation data are in excellent agreement with the theoretical result over the entire range of the rescaled variable, reproducing the main features predicted by our analysis: In particular, the distribution exhibits a logarithmic divergence at the origin, reflecting the enhanced statistical weight of trajectories associated with small instantaneous diffusivities.  At large values of 
$X/\sqrt{t}$, the PDF displays Gaussian tails modulated by a power-law dependence on $X$ with a non-universal exponent 
$2 \lambda \tau$. This behaviour is clearly captured by the simulations  -- the distribution becomes progressively narrower and more concentrated around the origin the larger $\lambda$ is -- which thus confirms the analytical prediction.  Physically, faster switching effectively averages the fluctuations of the stochastic diffusivity over shorter time scales, thereby reducing the magnitude of typical displacements.

\subsection{Long-time evolution of $P(X,t)$}

Consider the characteristic function of the particle displacement defined by Eq. \eqref{LE}, to find that for any fixed realization of the stochastic diffusivity $D_t$ one has
\begin{align}
	\label{phi}
	\Phi_{\omega} = \overline{\exp\left(\mathrm{i} \omega X_t\right)} = \exp\left(- \omega^2 \, {\cal D}_t \, t \right) \,,
	\end{align}
	with ${\cal D}_t$ being the time-averaged stochastic diffusivity,
	\begin{align}
		\label{d}
		{\cal D}_t = \frac{1}{t} \int^t_0 \mathrm{d}s \, D_t = \frac{1}{t} \int^t_0 \mathrm{d}s \, Y^2_s \,, 
		\end{align}
		where $Y_s$ obeys Eq. \eqref{dicho}. 
		Because the process 
	in Eq.~\eqref{dicho} is initialized at 
	$t= -\infty$,
	it is stationary for 
	$t > 0$. Consequently, 
	$\langle {\cal D}_t \rangle_{\xi} \equiv D_{\rm eff}$, with 
	$D_{\rm eff}$
	defined in Eq.~\eqref{z2}, as one may verify very directly. 
	
	In turn, we find that the variance of ${\cal D}_t$,  
\begin{align}
	\label{dd}
	{\rm Var}\left({\cal D}_t \right) = \langle {\cal D}^2_t \rangle_{\xi} - \langle {\cal D}_t \rangle^2_{\xi} \,,
	\end{align}
obeys, for  $t \to \infty$,  
\begin{align}
	\label{var1}
	{\rm Var}\left({\cal D}_t \right) =	\frac{4 \lambda A^4 \tau^5}{(1 + 2  \lambda \tau)^3} \frac{\tau}{t} + O\left(\frac{1}{t^2}\right) \,.
	\end{align}
	Expression for ${\rm Var}({\cal D}_t)$ valid for any $t$ is derived in \ref{D} and shows that the variance is a monotonically decreasing function of $t$. Note, as well, that in the diffusion limit the expression \eqref{var1} converges  to ${\rm  Var}({\cal D}_t) = \sigma^4 \tau^3/(2 t)$, as  it  should. 
	In \ref{D} we also present explicit expressions for the covariance function of ${\cal D}_t$.

Since ${\rm Var}\left({\cal D}_t \right) \to 0$ as $t \to \infty$, the random variable ${\cal D}_t$ converges in mean square to its mean value given by Eq. \eqref{z2}. In other words, 
${\cal D}_t$
is a self-averaging quantity. The relative magnitude of its fluctuations around the mean decreases with time, with deviations scaling as 
$1/\sqrt{t}$. As a result, temporal fluctuations of the stochastic diffusivity become progressively less important on long time scales, and the dynamics is effectively governed by the mean diffusivity. It is therefore natural to expect that the position PDF converges to a Gaussian form in the long-time limit.

To demonstrate the convergence to a Gaussian distribution and to estimate the rate of this convergence, we construct the Edgeworth  expansion \cite{fellerIntroductionProbabilityTheory1968}. Following this standard approach, we analyse the cumulant generating function and expand it, retaining the two leading contributions 
\begin{align}
	\ln \langle\Phi(\omega)\rangle_{\xi}=
	-\omega^2 t \langle {\cal D}_t\rangle_{\xi}
	+\frac{\omega^4 t^2}{2}{\rm Var}({\cal D}_t)
	+O(\omega^6) \,,
\end{align}
where $\Phi(\omega)$ is defined in Eq. \eqref{phi}.
Within this approximation, the averaged characteristic function reads
\begin{equation}
	\langle \Phi(\omega) \rangle_{\xi}  \simeq
	\mathrm{e}^{-\omega^2 D_{\rm eff} t }
	\left[1+\frac{\omega^4 t^2}{2}{\rm Var}({\cal D}_t)\right] \,.
\end{equation}
The Fourier transform of the above expression yields the two leading terms of the (infinite) Edgeworth series for the position PDF:
\begin{align}
		\label{eq:LPDX}
		P(X,t) \simeq \frac{1}{\sqrt{4\pi D_{\rm eff} t}}
		\exp\!\left(-\frac{X^2}{4 D_{\rm eff} t}\right)
		\left[
		1+
		\frac{\mathrm{Var}({\cal D}_t)}{8 D_{\rm eff}^2}
		H_4\!\left(\frac{X}{\sqrt{2 D_{\rm eff} t}}\right)
		\right] \,,
\end{align}
where 
$H_4(z)=z^4-6z^2+3$ is the fourth Hermite polynomial and $D_{\rm eff}$ is defined in Eq. \eqref{z2}.
Since ${\rm Var}({\cal D}_t) \sim 1/t$, the correction vanishes at long times and the distribution converges to a Gaussian function. Note, as well, that in the diffusion limit (see  Eq. \eqref{limit}) the expression \eqref{eq:LPDX} reduces to the result obtained in \cite{chechkinBrownianNonGaussianDiffusion2017,lanoiseleeModelNonGaussianDiffusion2018a}. Higher-order contributions to the Edgeworth expansion can, in principle, be derived using the same procedure, but are not considered here.

\begin{figure}[t]
	\includegraphics[width=0.31\textwidth]{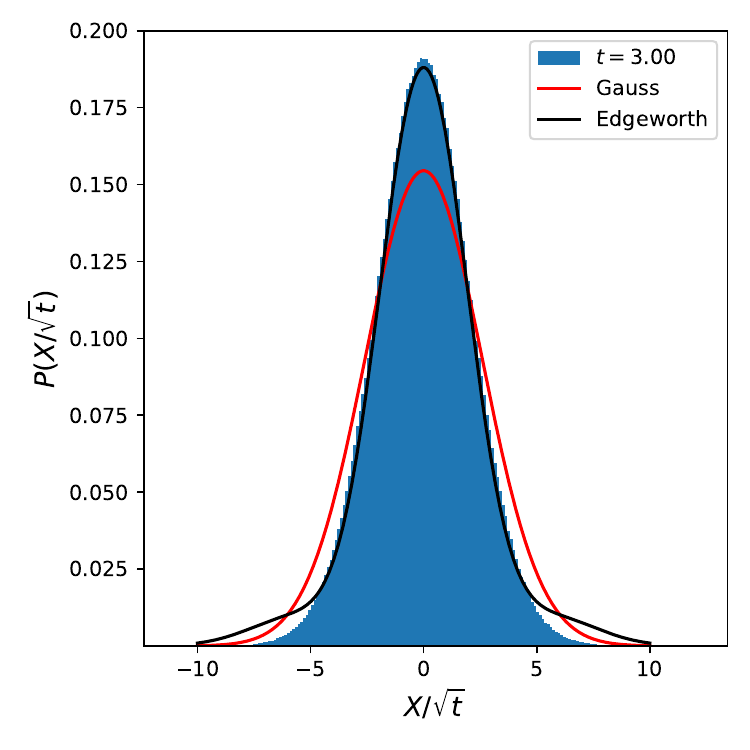}
\includegraphics[width=0.31\textwidth]{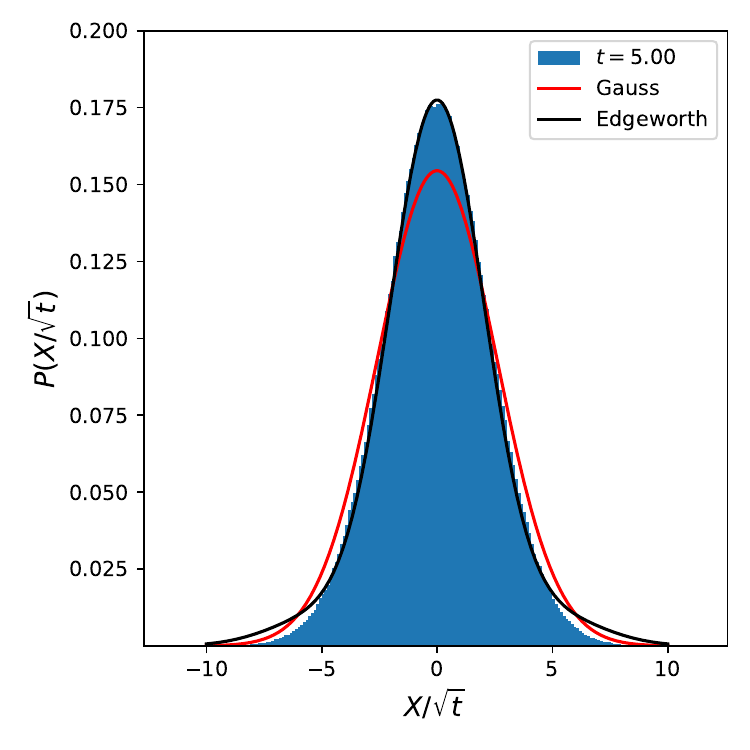}
\includegraphics[width=0.31\textwidth]{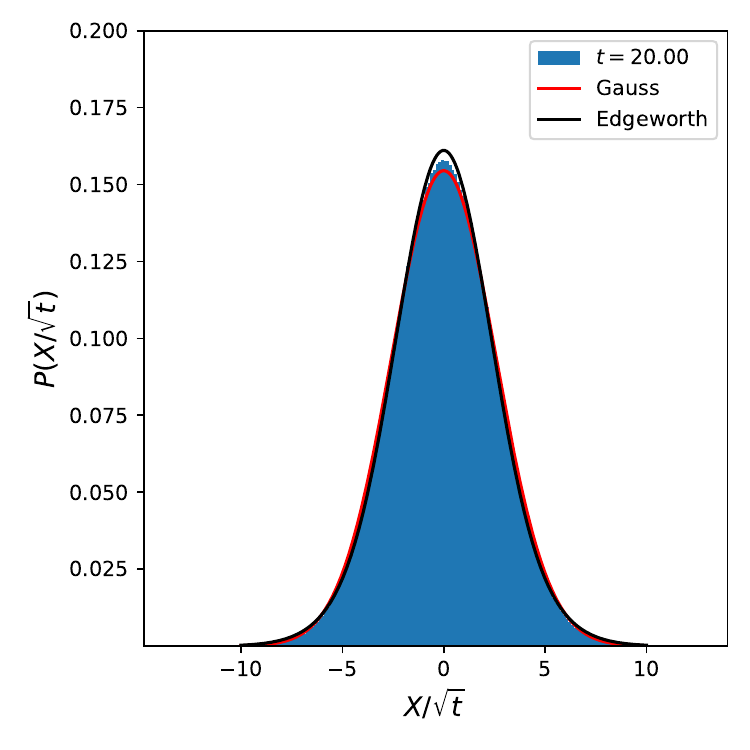}
	\caption{The position PDF $P(X,t)$ in the DD model driven by dichotomous noise for $\tau = 1$, $A  = 1$, $\lambda=0.1$  and three moderately short values of time $t$ :
		$t = 3 $ (left),  $t = 5$ (center)  and $t = 20$ (right). The thin black curves correspond to the theoretical prediction in Eq.~\eqref{eq:LPDX}, including the first two terms of the Edgeworth expansion, while the red curve represents the leading Gaussian contribution only. The histograms depict our numerical simulations results.}\label{fig:PdeXmiddle}
\end{figure}
We now examine the convergence of the position PDF to the Gaussian limit and assess the relevance of the second term in the Edgeworth expansion. To this end, we focus on moderately short times ($t=3,5$ and $20$)
 and consider a small value of the switching rate, 
$\lambda = 0.1$, for which deviations from Gaussian behaviour are expected to be most pronounced.
In Fig.~\ref{fig:PdeXmiddle}, we present the position PDF 
$P(X,t)$, plotted as a function of the rescaled variable 
$X/\sqrt{t}$. The numerical results (blue histograms) are compared with the two-term Edgeworth approximation given by Eq.~\eqref{eq:LPDX} (black curves), as well as with the leading Gaussian contribution alone (red curves). For 
$t=3$, the Gaussian approximation exhibits substantial deviations from the numerical data, failing to capture both the central behaviour and the overall shape of the distribution. In contrast, the two-term Edgeworth expansion provides a significantly improved description, accurately reproducing the central part of the PDF even for such a small value of $t$. At 
$t=5$, the Gaussian approximation performs better but still remains quantitatively inaccurate, whereas the Edgeworth approximation continues to yield an excellent description of the central region. Upon further increasing the time to 
$t=20$, both approximations become nearly indistinguishable and closely match the numerical results, with only minor residual discrepancies in the central part. These observations demonstrate that the second term in the Edgeworth expansion plays a crucial role in capturing non-Gaussian features at moderately short times, while its contribution progressively diminishes as the system approaches the asymptotic Gaussian regime.

\begin{figure}[H]
	\includegraphics[width=0.31\textwidth]{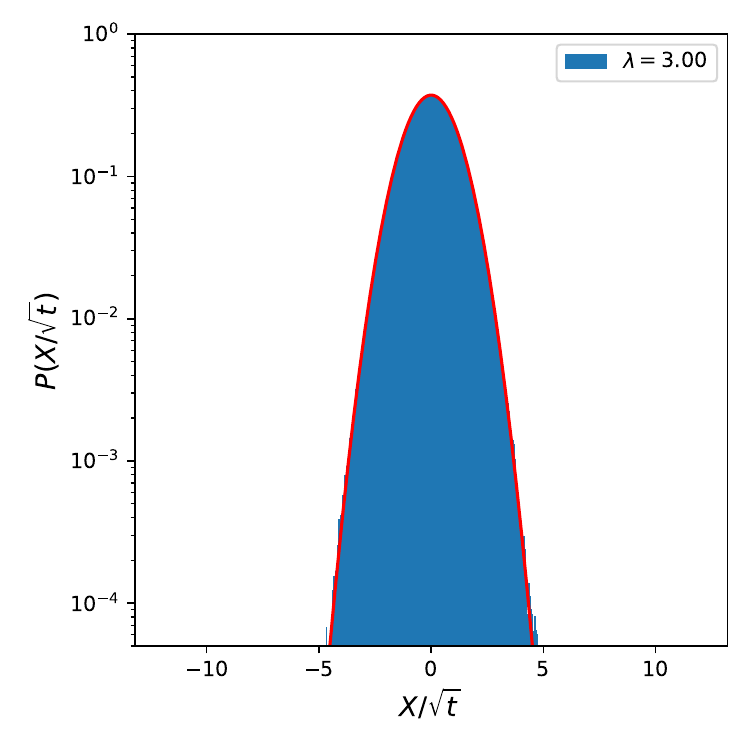}
\includegraphics[width=0.31\textwidth]{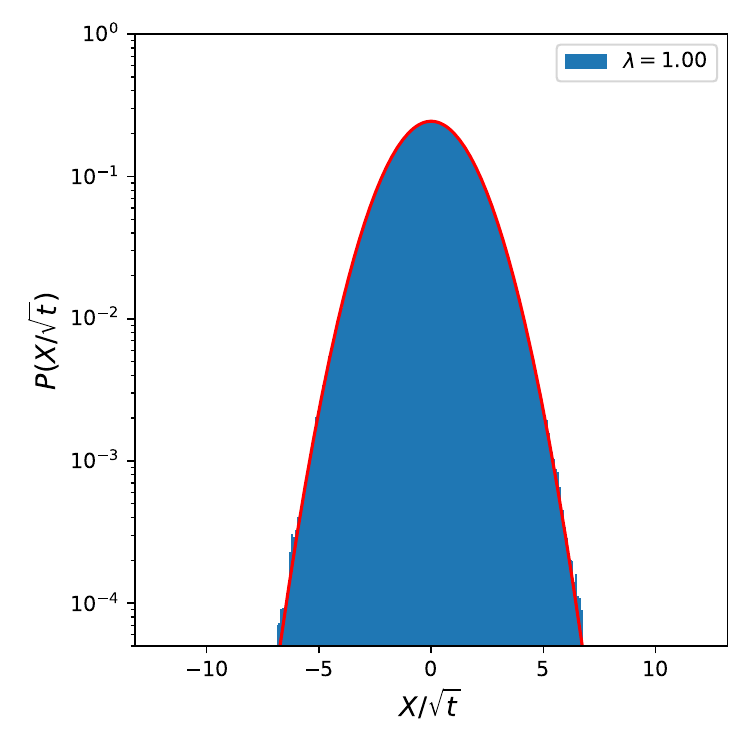}
\includegraphics[width=0.31\textwidth]{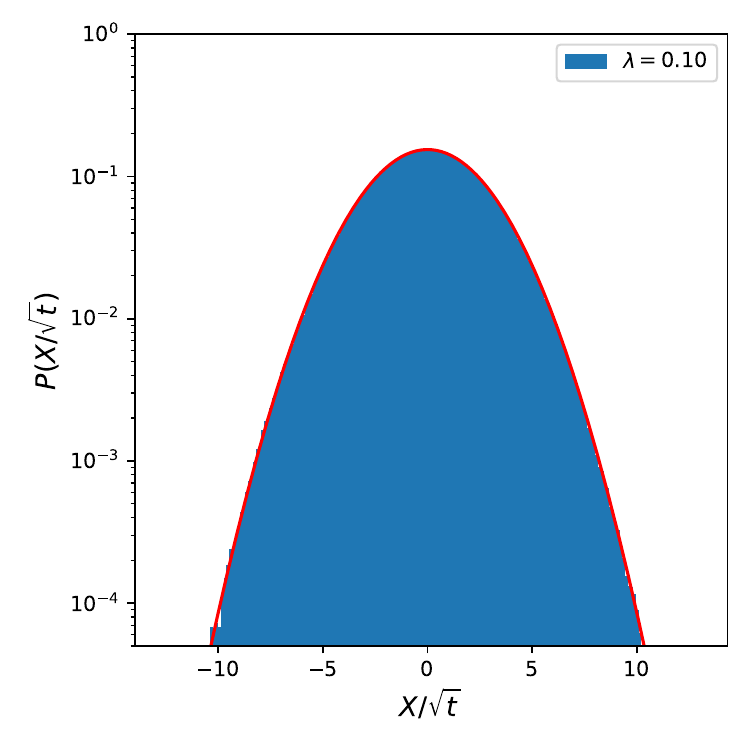}
	\caption{The position PDF $P(X,t)$ in the DD model driven by dichotomous noise for $\tau = 1$, $A  = 1$, $t = 100$  and three values of the switching rate $\lambda$: 
		$\lambda=3$ (left), $\lambda = 1$ (center) and $\lambda = 0.1$ (right). Thin red curves correspond to our theoretical prediction in Eq. \eqref{eq:LPDX}. The blue histograms depict our numerical simulations results.}\label{fig:PdeX}
\end{figure}

We next examine the behaviour of the position PDF at sufficiently large times, focusing on its dependence on the switching rate 
$\lambda$. Figure~\ref{fig:PdeX} shows 
$P(X,t)$
for 
$t=100$, plotted as a function of the rescaled variable 
$X/\sqrt{t}$, for several values of 
$\lambda$. The histograms correspond to numerical simulations (see \ref{A}), while the solid red curves represent the analytical prediction given by the first term in Eq.~\eqref{eq:LPDX}. We observe that at such long times, all distributions closely approach a Gaussian form for all considered values of 
$\lambda$,  indicating that higher-order corrections are negligible in this regime. Despite this universal Gaussian shape, the width of the distributions retains a clear dependence on the switching rate. In particular, the distribution is broadest for 
$\lambda = 0.1$ and becomes progressively narrower as 
$\lambda$ increases, with the most narrow profile observed for 
$\lambda = 3$. This reflects the fact that faster switching effectively reduces the magnitude of fluctuations by averaging the stochastic diffusivity over shorter time scales.
Overall, these results demonstrate that while the long-time behaviour is universally Gaussian, the switching rate 
$\lambda$ controls the effective diffusivity and hence the width of the asymptotic distribution.

\section{Conclusions}
\label{concs}

We studied a variant of the DD model in which the OU process governing the stochastic diffusivity is driven by a symmetric dichotomous (telegraph) noise. This modification preserves the mean-reverting dynamics of the original formulation but replaces Gaussian environmental fluctuations by bounded switching dynamics. As a consequence, the stationary diffusivity is confined to a finite interval, in contrast to the unbounded gamma distribution obtained in the Gaussian OU model. The shape of the distribution for the dichotomous OU case depends on the switching rate:  for sufficiently large 
$\lambda$ it can be a monotonic function of $D$, with a singularity at the left edge of the support , 
or for sufficiently small values of $\lambda$ be  a
$U$-shaped function with integrable singularities at both boundaries.

We showed that this modification leads to some  similarities but also to quantitative differences in the displacement statistics in the short-time regime. In particular, analytical expressions for the short-time position probability density function were obtained in terms of the confluent hypergeometric (Tricomi) function. As in the Gaussian OU model, the distribution exhibits a logarithmic divergence at the origin due to trajectories with very small instantaneous diffusivity. The behaviour of the tails, however, differs qualitatively. While the Gaussian OU model yields exponential tails at short times, the bounded diffusivity in the dichotomous OU case leads to Gaussian tails divided by a power-law factor with a non-universal exponent dependent on the switching rate and the relaxation time. These results illustrate the role of environmental fluctuations in determining the non-Gaussian features of stochastic transport.
Despite these differences at short and intermediate times, both models exhibit the same asymptotic behaviour. Because the stochastic diffusivity has a finite correlation time, the displacement statistics converge at long times to ordinary Gaussian diffusion with an effective diffusion coefficient determined by the stationary statistics of the diffusivity.

More generally, the model provides a minimal analytically tractable framework for stochastic transport in environments where fluctuations occur through intermittent switching between discrete dynamical states. Such situations arise in heterogeneous or compartmentalized media, in active systems with alternating activity states, and in stochastic models with regime-switching parameters. The telegraph-driven formulation thus complements the standard Gaussian DD model and highlights the impact of bounded environmental fluctuations on diffusion statistics.
Possible extensions include the analysis of first-passage and extreme-value properties, higher-dimensional generalizations, and more complex switching dynamics or multi-state environments.

\section*{Acknowledgments}
J.-H.J. gratefully acknowledges financial support from the Department of Physics of Sorbonne University and thanks the Laboratory for Theoretical Condensed Matter Physics (Sorbonne University/CNRS) for its warm hospitality during his visit in January–February 2026, when this work was initiated. The National Research Foundation of Korea, Grants No. RS-2024-00343900, is acknowledged (J.-H.J).
\section*{References}

\bibliography{refs2.bib}

\appendix

\section{Numerical simulations}
\label{A}

In this appendix, we describe the numerical procedures used to simulate the stochastic dynamics and to evaluate the statistical properties of the position probability density function.
Numerical simulations are performed using the Euler–Maruyama integration scheme for stochastic differential equations~\cite{kloedenNumericalSolutionStochastic1992,highamAlgorithmicIntroductionNumerical2001}. This method provides a straightforward and reliable discretization for Langevin-type dynamics driven by stochastic processes.

To analyse the short-time behaviour of the position PDF $P(X,t)$, it is essential to ensure that the stochastic diffusivity is initially distributed according to its stationary state. To this end, the instantaneous diffusion coefficient is sampled directly from its stationary distribution given in Eq.~\eqref{eq:sanchobis}. This procedure avoids long equilibration transients and ensures that the process is stationary for all $t>0$, consistently with the theoretical framework in which the auxiliary process $Y_t$ is assumed to evolve from $t=-\infty$.
For convenience, the diffusion coefficient is rescaled by the factor $A^2\tau^2$, which confines its support to the unit interval $[0,1]$. The stationary distribution of the rescaled diffusivity $D$ is then given by
\begin{align}\label{eq:sanchoter}
	P_{\rm st}(D) =
	\frac{\Gamma(\lambda \tau + \tfrac{1}{2})}{\Gamma(\lambda \tau)\sqrt{\pi D}}
	(1-D)^{\lambda \tau - 1}, \quad 0 \leq D \leq 1.
\end{align}
This beta distribution exhibits an integrable singularity at the origin. Its behaviour near $D = 1$ depends on the parameter $\lambda\tau$: it vanishes for $\lambda\tau>1$, diverges for $\lambda\tau<1$, and remains finite in the marginal case $\lambda\tau=1$.

The corresponding cumulative distribution function reads
\begin{align}
	F(D) = \int_{0}^{D} P_{\rm st}(x)\,\mathrm{d}x
	= \frac{2\sqrt{D}\,\Gamma\!\left(\lambda\tau+\tfrac{1}{2}\right)
		{}_2F_1\!\left(\tfrac{1}{2},1-\lambda\tau;\tfrac{3}{2};D\right)}
	{\sqrt{\pi}\Gamma(\lambda\tau)},
\end{align}
where ${}_2F_1$ denotes the Gauss hypergeometric function~\cite{batemanHigherTranscendentalFunctions1955}.
Random realizations of the diffusivity are generated using the inverse transform sampling method~\cite{devroyeNonUniformRandomVariate1986}. Specifically, independent random numbers $\eta$ are drawn from a uniform distribution on $[0,1)$, and the corresponding values of $D$ are obtained by solving the equation
\begin{equation}
	F(D) = \eta.
\end{equation}
The solution is found numerically in the interval $[0,1)$ using Brent’s root-finding algorithm~\cite{brentAlgorithmsMinimizationDerivatives2013,pressNumericalRecipes3rd2007}, which combines bisection, secant, and inverse quadratic interpolation methods and ensures both robustness and rapid convergence.

The Langevin equation governing the particle position is integrated with a fixed time step $\delta t = 0.005$. The parameters are set to $\tau = 1$ and $A = 1$. The dichotomous noise is generated as a telegraph process with switching rate $\lambda$ by sampling exponentially distributed waiting times between successive switches.
The simulations are initialized at $t=0$. To reproduce the stationary regime corresponding to an infinitely long prehistory, the initial value $Y_0$ of the auxiliary OU process is drawn from its equilibrium distribution. This ensures that the stochastic diffusivity is stationary from the outset, in agreement with the analytical assumptions.

Ensemble averages are computed over $n = 10^5$–$10^6$ independent trajectories, depending on the quantity of interest. This large number of realizations ensures good statistical convergence of the measured observables, including the full shape of the probability density function and its tails. In particular, careful sampling is required to accurately resolve non-Gaussian features and rare fluctuations.

We have verified that the numerical results are robust with respect to the choice of the time step and the sampling procedure. In all cases, excellent agreement is observed between the simulation data and the analytical predictions presented in the main text.

\section{Covariances of the squared dichotomous OU process $Y_t^2$ and of the time-averaged stochastic diffusivity ${\cal D}_t$ }
\label{D}

Consider the process defined in Eq.~\eqref{dicho} and introduce the centered variables
\begin{equation}
	X_1(t) = Y_t^2 - \langle Y_t^2 \rangle_{\xi}, 
	\quad 
	X_2(t) = \xi_t Y_t - \langle \xi_t Y_t \rangle_{\xi}.
\end{equation}
Since $Y_t$ is stationary (it is defined with initial time $t=-\infty$), its second moment is time-independent and given by $\langle Y_t^2 \rangle_{\xi} = D_{\rm eff} = A^2 \tau^2/(1+2\lambda\tau)$.

\subsection{Covariance function of the squared dichotomous OU process $Y_t^2$}

We aim to determine the covariance function
\begin{equation}
	C_X(t) = \langle X_1(t) X_1(0) \rangle_{\xi}
	= \langle Y_t^2 Y_0^2 \rangle_{\xi} - \langle Y_t^2 \rangle_{\xi}^2.
\end{equation}
Once $C_X(t)$ is known, it can be used to evaluate the variance and covariance of the time-averaged stochastic diffusivity ${\cal D}_t$.

Multiplying Eq.~\eqref{dicho} by $2Y_t$ and centering the result yields
\begin{equation}
	\frac{\mathrm{d}}{\mathrm{d}t} X_1(t) = -\frac{2}{\tau} X_1(t) + 2A X_2(t),
\end{equation}
where constant terms cancel due to stationarity. Multiplying by $X_1(0)$ and averaging, we obtain
\begin{equation}
	\label{1st_short2}
	\frac{\mathrm{d}}{\mathrm{d}t} C_X(t) = -\frac{2}{\tau} C_X(t) + 2A \langle X_2(t) X_1(0) \rangle_{\xi}.
\end{equation}

The equation is not closed, as it involves the mixed correlation $\langle X_2(t) X_1(0) \rangle_{\xi}$. To determine its evolution, we differentiate it with respect to time and use the definition of $X_2(t)$:
\begin{equation}
	\frac{\mathrm{d}}{\mathrm{d}t} \langle X_2(t) X_1(0)\rangle_{\xi}
	= \langle \dot{\xi}_t Y_t X_1(0)\rangle_{\xi}
	+ \langle \xi_t \dot{Y}_t X_1(0)\rangle_{\xi}.
\end{equation}
The first term is treated using the Shapiro–Loginov identity~\cite{shapiroFormulaeDifferentiationTheir1978}, while the second follows directly from the Langevin equation. After straightforward algebra, we obtain
\begin{equation}
	\label{2nd_short2}
	\frac{\mathrm{d}}{\mathrm{d}t} \langle X_2(t) X_1(0) \rangle_{\xi}
	= -\left(2\lambda + \frac{1}{\tau}\right) \langle X_2(t) X_1(0) \rangle_{\xi}.
\end{equation}

Equations~\eqref{1st_short2} and \eqref{2nd_short2} form a closed system of linear differential equations. The initial conditions are obtained using the stationarity of the process $Y_t^2$. We have
\begin{align}
A \tau  \langle \xi_t Y_t  \rangle_{\xi} = \langle Y^2_t  \rangle_{\xi} \,, \quad	A \tau \langle \xi  Y^3 \rangle_{\xi} =  	\langle Y^4 \rangle_{\xi} \,,
\end{align}
while the fourth moment of $Y_t$ obtains from Eq. \eqref{mom}:  
\begin{align}
	\langle Y^4  \rangle_{\xi} = \frac{3 A^4 \tau^4}{(1 + 2 \lambda \tau)(3  + 2 \lambda \tau)} \,.
	\end{align}
Combining the above expressions, we find
\begin{equation}
	C_X(0) = \mathrm{Var}(Y^2),
	\quad 
	\langle X_2(0) X_1(0) \rangle_{\xi} = \frac{\mathrm{Var}(Y^2)}{A\tau},
\end{equation}
where
\begin{equation}
	\mathrm{Var}(Y^2) = \frac{4\lambda\tau}{(1+2\lambda\tau)^2(3+2\lambda\tau)} A^4 \tau^4.
\end{equation}

\begin{figure}[t]
	\includegraphics[width=0.95\textwidth]{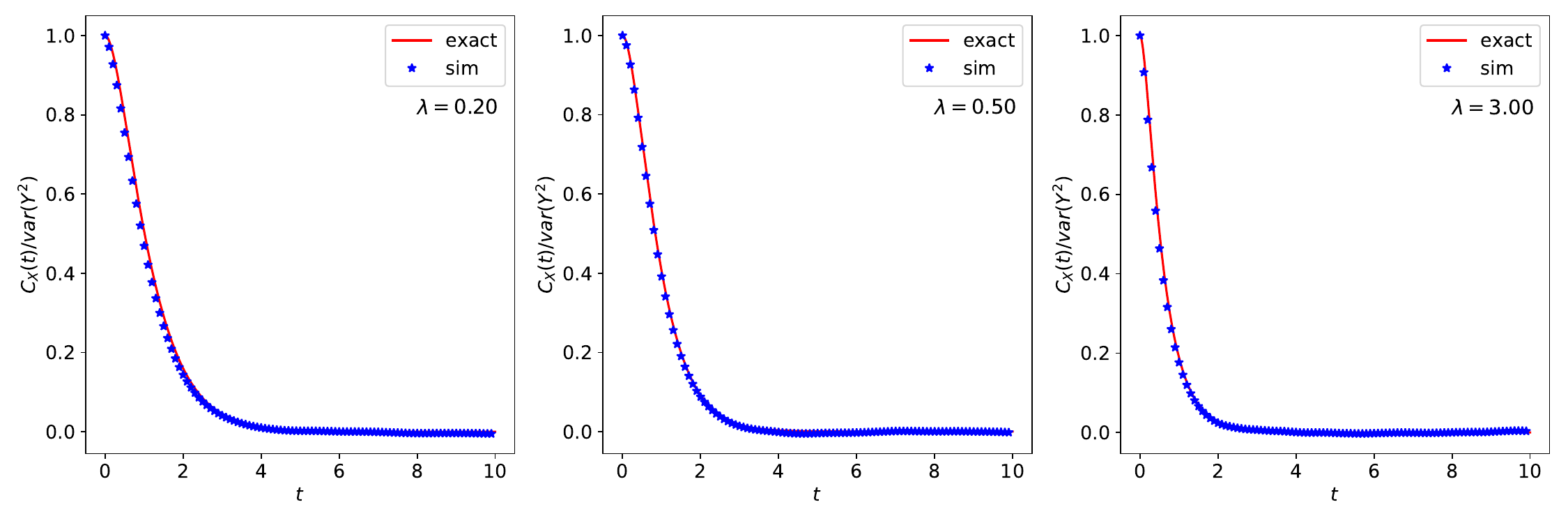}
			\caption{Reduced covariance function $C_X(t)/{\rm Var}(Y^2)$ as function of time for three values of the switching rate $\lambda$: $\lambda = 0.2$ (left), $\lambda = 0.5$ (center) and $\lambda = 3$ (right). The relaxation time $\tau = 1$. Thin red curves correspond to our Eqs. \eqref{CX_final2} and \eqref{Cpart}. Symbols depict the results of numerical simulations.}\label{fig:5}
\end{figure}

Solving the system of differential Eqs. \eqref{1st_short2} and \eqref{2nd_short2}, we find that 
\begin{equation}
	\label{CX_final2}
	C_X(t) = \frac{\mathrm{Var}(Y^2)}{1-2\lambda\tau}
	\left[
	2 \mathrm{e}^{-(1+ 2 \lambda \tau)t/\tau} - (1+2\lambda\tau)\mathrm{e}^{- 2 t/\tau}
	\right] \,, \quad 2\lambda\tau \neq 1 \,.
\end{equation}
 The covariance consists of two exponential contributions with different decay rates and  amplitudes with opposite signs. Nonetheless,  
$C_X(t)$ is always positive and vanishes when $t \to \infty$. 

The case $2\lambda\tau = 1$ is special because the two decay rates coincide and therefore this case should be considered separately. Solving the system of differential Eqs. \eqref{1st_short2} and \eqref{2nd_short2} with $2 \lambda \tau$ set equal to $1$, we find  
that here the solution attains the form
\begin{equation}
	\label{Cpart}
	C_X(t) = \frac{A^4 \tau^4}{8}\left(1+\frac{2t}{\tau}\right)\mathrm{e}^{-2t/\tau} \,, \quad 2\lambda\tau = 1 \,.
\end{equation}
In this case as well, $C_X(t)$ is also manifestly positive. 

Our analytical predictions for the covariance function of the squared dichotomous OU process are shown in Fig.~\ref{fig:5}, together with numerical results obtained for relaxation time $\tau = 1$ and three values of the switching rate: $\lambda = 0.2$ and $\lambda = 3$ (left and right panels, corresponding to Eq.~\eqref{CX_final2}), and $\lambda = 0.5$ (central panel, the critical case described by Eq.~\eqref{Cpart}). We observe excellent agreement between the analytical expressions and the numerical data.

\subsection{Variance of the time-averaged stochastic diffusivity}

Consider next the variance ${\rm Var}({\cal D}_t)$ of the time-averaged stochastic diffusivity ${\cal D}_t$, Eq. \eqref{d}, which is defined in Eq. \eqref{dd}. Since the process $Y_t$ is stationary, we have
\begin{align}
	\begin{split}
	{\rm Var}({\cal D}_t) &= \frac{1}{t^2} \int^t_0 \int^t_0 \mathrm{d}s \, \mathrm{d}s' \, \left[\langle Y_s^2 Y_{s'}^2\rangle - \langle Y_s^2\rangle \langle Y^2_{s'}\rangle\right]\\
	&=\frac{1}{t^2} \int^t_0 \int^t_0 \mathrm{d}s \, \mathrm{d}s' \, C_X(|s - s'|) = \frac{2}{t^2} \int^t_0 \mathrm{d}s\,(t - s) C_X(s) \,.
	\end{split}
	\end{align}
Performing the integral in the latter expression, we obtain the following result, valid for any  $t$ and arbitrary value of  $2 \lambda \tau$ (including $2 \lambda \tau = 1$),  
	\begin{align}
		\begin{split}
			\label{d3}
		{\rm Var}({\cal D}_t) 	= \frac{ (3 + 2 \lambda \tau) \, {\rm Var}(Y^2) \, \tau}{(1 + 2 \lambda \tau) \, t}
			&\Bigg[1  - \frac{4 \tau}{(1 - (2 \lambda \tau)^2)(3 + 2 \lambda \tau) \, t} \left(1 - \mathrm{e}^{-(1 + 2 \lambda \tau) t/\tau}\right)
			\\
			&+ \frac{(1 + 2 \lambda \tau)^2 \tau}{2 (1 - 2 \lambda \tau) (3 + 2 \lambda \tau) \, t} \left(1 - \mathrm{e}^{- 2 t/\tau}\right)
			\Bigg] \,.
			\end{split}
		\end{align}
		Note that ${\rm Var}({\cal D}_t) = {\rm Var}(Y^2)$ for $t = 0$ and is a monotonically decreasing function of time, ${\rm Var}({\cal D}_t) \propto 1/t$, which signifies that the time-averaged stochastic diffusivity ${\cal D}_t$ is self-averaging. The leading large-$t$ term is presented in Eq. \eqref{var1} in the main text.
\begin{figure}[t]
	\includegraphics[width=0.95\textwidth]{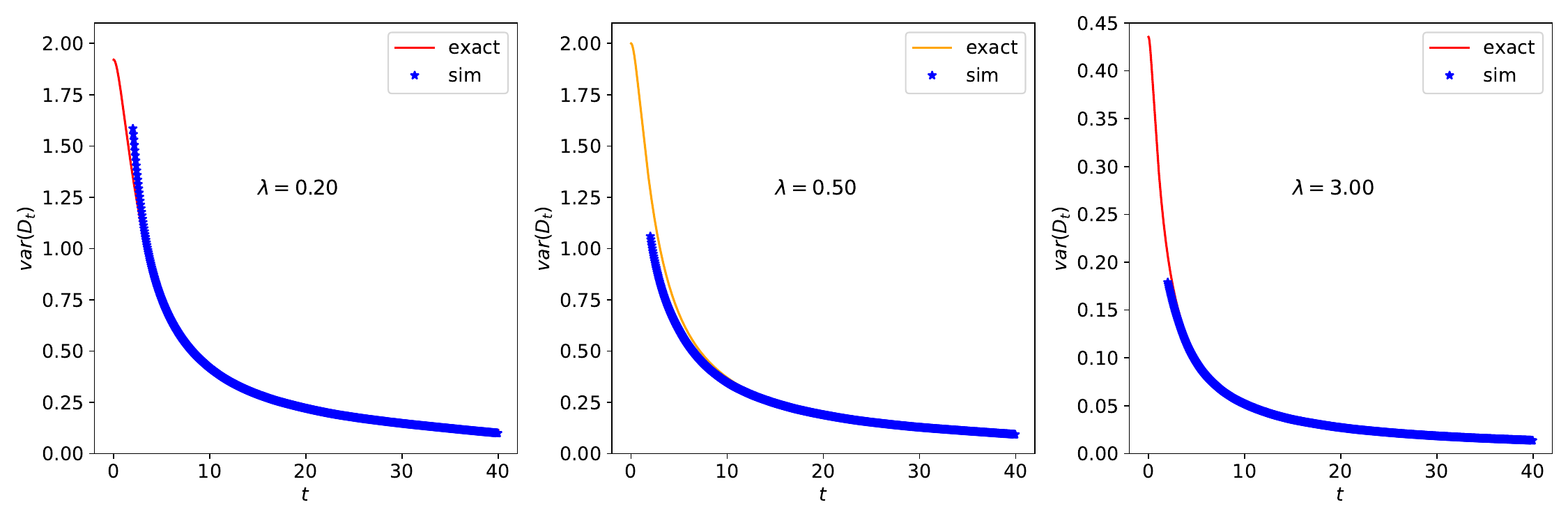}
	\caption{Variance ${\rm Var}({\cal D}_t)$ of the time-averaged stochastic diffusivity as function of time for three values of the switching rate $\lambda$: $\lambda = 0.2$ (left), $\lambda = 0.5$ (center) and $\lambda = 3$ (right). The relaxation time $\tau = 1$. Thin red curves correspond to our Eq. \eqref{d3}. Symbols depict the results of numerical simulations for $t  > 5$.}\label{fig:6}
\end{figure}
Figure~\ref{fig:6} presents our analytical results for the variance of the time-averaged stochastic diffusivity associated with the squared dichotomous OU process, alongside numerical simulations. The data are shown for a relaxation time $\tau = 1$ and three values of the switching rate: $\lambda = 0.2$ and $\lambda = 3$ (left and right panels), and $\lambda = 0.5$ (central panel). Thin red curves are our analytical predictions in Eq.  \eqref{d3}, while the symbols depict the numerical results obtained for sufficiently large values of  $t$, ($t > 5$). At small times 
$t$, numerical analysis becomes prohibitively involved, owing to the requirement of a highly refined temporal discretisation with very small time steps.

\subsection{Covariance of the time-averaged stochastic diffusivity}  

Here we calculate the covariance function of the  time-averaged  stochastic diffusivity defined as
\begin{align}
	\begin{split}
	{\rm  Cov}_{\cal D}(t,t') &= \langle {\cal D}_t {\cal D}_{t'} \rangle -  \langle {\cal D}_t \rangle \langle {\cal D}_{t'} \rangle \\
	& = \frac{1}{t \, t'} \int^t_0 \int^{t'}_0 \mathrm{d}s \, \mathrm{d}s' \, C_X(|s -s'|) \,.
	\end{split}
	\end{align}
	Inserting the expression \eqref{CX_final2} (or \eqref{Cpart}) into the above definition and performing integrations, we find that for $t \leq t'$, the covariance function of the time-averaged stochastic diffusivity obeys, for any value of $2 \lambda \tau$ (including $2 \lambda \tau = 1$)
\begin{align}	
	\begin{split}
	\label{CovD}
	{\rm Cov}_{\cal D}(t,t')
&= \frac{t}{t'} {\rm Var}({\cal D}_t)
+ \frac{{\rm Var}(Y^2)}{(1-2\lambda\tau) \, t \, t'} \\&\times\Bigg[2 \, \Delta_{t,t'}\left(\alpha=\frac{(1 + 2 \lambda \tau)}{2}\right) - (1 + 2 \lambda \tau) \, \Delta_{t,t'}\left(\alpha = \frac{2}{\tau}\right)\Bigg]\,, \quad t \leq t'\,, 
\end{split}
\end{align}
where ${\rm Var}({\cal D}_t)$ is defined in Eq. \eqref{d3},  while 
\begin{align}
	\begin{split}
		\label{delta}
\Delta_{t,t'}(\alpha) =\frac{1}{\alpha^2} \Bigg[\mathrm{e}^{-\alpha t} + \mathrm{e}^{-\alpha t'} - \mathrm{e}^{-\alpha (t' - t)} - 1\Bigg] \,, \quad t \leq t' \,.
\end{split}
\end{align}
For $t \geq t'$, ${\rm Cov}_{\cal D}(t,t')$ is obtained from Eqs.~\eqref{CovD} and \eqref{delta}  by swapping $t$ and $t'$. 

The covariance structure of the time-averaged diffusivity ${\cal D}_t$ reflects both the finite-time averaging and the two-mode relaxation encoded in $C_{X}(t)$. In the equal-time case $t = t'$, the covariance reduces to the variance ${\rm Var}({\cal D}_t)$, which we computed independently in the previous subsection. At short times $t \ll \tau$, one finds ${\rm Var}({\cal D}_t) \to {\rm Var}(Y^2)$, as expected: in this case the averaging window is too short to reduce fluctuations, so ${\cal D}_t$ essentially samples the instantaneous variable $Y_t^2$. As time increases, the variance decreases due to temporal averaging. For $t \gg \tau$, the exponentials in the exact expression vanish and the leading behavior becomes
\begin{align}
{\rm Var}({\cal D}_t) \sim \frac{{\rm Var}(Y^2)\,\tau}{t} 
\end{align}
demonstrating self-averaging: fluctuations of the time-averaged diffusivity decay inversely with the observation time.

For unequal times, the covariance ${\rm Cov}({\cal D}_t,{\cal D}_{t'})$ (assuming $t \le t'$) naturally splits into a dominant contribution proportional to $t {\rm Var}({\cal D}_t)/t'$, reflecting the overlap of the two averaging intervals, plus a correction term involving exponentials. When $t'$ becomes much larger than $t$, this prefactor $t/t'$ drives the covariance to zero, expressing the loss of correlation between averages taken over vastly different time windows. The exponential correction terms decay on the scale $\tau$ and become negligible for $t, t' \gg \tau$, so that the covariance asymptotically behaves as
\begin{align}
{\rm Cov}({\cal D}_t,{\cal D}_{t'}) \sim \frac{t}{t'} {\rm Var}({\cal D}_t),
\end{align}
which itself vanishes as $1/t'$. Thus, not only does each ${\cal D}_t$ self-average, but time-averaged observables over increasingly separated intervals become asymptotically uncorrelated. This behavior is a direct consequence of the finite correlation time of $Y_t^2$ combined with the growing averaging window.

\end{document}